Correlative statistical microstructural assessment of precipitates and their distribution, with simultaneous electron backscatter diffraction and energy dispersive X-ray spectroscopy

Chris Bilsland*[1], Andrew Barrow[2], and T. Ben Britton[1]

1. Department of Materials, Imperial College London, SW7 2AZ
2. Rolls-Royce Submarines Limited, PO Box 2000, Raynesway, Derby, DE21 7XX

*Corresponding author: C.Bilsland17@imperial.ac.uk

**Abstract**

Modern engineering alloys have bespoke microstructures, where features such as precipitates are used to control properties. In many Ni-based alloys, carbo-nitride precipitates are introduced to strengthen and improve performance. These precipitates can be distributed throughout the microstructure and niobium rich carbides are often found at grain boundaries. In this work, we used combined energy dispersive X-ray spectroscopy (EDS) and electron backscatter diffraction (EBSD) to characterise the population of these precipitates. Processing of the EDS signal is used to label the Mo/Nb-rich precipitates, and their size and location are measured from maps using a circular Hough transform. This label map is combined with the grain boundary network (from EBSD analysis). Statistical analysis, using ANOVA testing, reveals differences in chemistry between carbides found in Ni-rich matrix grain interiors, on random high angle boundaries and on special boundaries ($\Sigma 3$ and $\Sigma 9$). These results are compared between wrought and power metallurgy product forms. These distributions are discussed in the context of their performance within demanding environments, such as reactor core internals.

**Introduction**

Engineering alloys play a significant role in enabling modern technology, especially in extreme environments such as found in nuclear reactor components. The performance and reliability of these alloys is informed through understanding of their microstructure and the impact of microstructural units in challenging environments leading to failure processes such as stress corrosion cracking. This motivates our present study to provide quantitative microstructural assessment, underpinned with statistical testing, of different microstructural units using electron backscatter diffraction (EBSD) and energy dispersive X-ray spectroscopy (EDS).

The diffraction patterns analysed by EBSD provides access to the structural information, e.g. phase classification and crystal orientations, and analysis of EBSD maps enable creation of grain boundary network maps where the different character of each grain boundary can be probed (such as the grain boundary misorientation and the surface trace of the boundary).

The X-ray emission spectra analysed by EDS, created by electron induced electron transitions of the core electrons of the probe volume, can be characterised to reveal variations in local chemistry and also assist in phase classification.

Both EDS and EBSD are powerful techniques individually when characterising materials. In recent years is becoming possible to use the data collected from these methods to provide even more information than is available to either technique individually, linking chemistry and structure directly and across large areas. Previous work has used simultaneous EBSD-EDS in a range of applications involving enhancing microstructural characterisation [1]–[8] and also efforts to create quantitative metrics for microstructural features [9]. What these methods have focussed on is the enhancement of the phase classification ability of EBSD with additional EDS processing to complement the categorisation. Other methods of enhancing EDS processing have included the development of X-ray image segregation through X-ray scatter diagrams by Moran and Wurher to find methods to segment X-ray maps based on EDS data [10], [11]. For the present work, we aim to combine the data more directly in a correlative approach, and providing information that would not usually be collected via different techniques.

The physics of electron scattering, X-ray generation and emission and electron diffraction mean that the interaction volumes for the techniques can be substantively different [12], [13]. This is very important to consider as usually combined EBSD-EDS is performed at high sample tilt and with high probe energies (20 keV). In the present work, we resolve this issue with Monte Carlo based modelling [14] to assist interpretation.

Beyond the physics, the signals and analysis methods for EDS and EBSD are also significantly different. EDS is a spectroscopy technique, and the signal obtained is a function of the electron scattering, X-ray generation, X-ray absorption and fluorescence combined with the detector position and type [15]–[17]. Often each spectrum can be processed by the commercial software for semi-quantitative analysis. In contrast, EBSD based Kikuchi patterns are typically indexed and classified by crystal structure and crystal orientation only and the chemical information cannot be extracted [18]. Using these methods for correlative microscopy requires combining of the two data modalities, and the pipeline may vary depending on the ultimate goal of the materials characterisation approach.

In this work, we aim to use simultaneous EBSD-EDS to quantitatively probe precipitate distributions in a Ni-metal matrix. To do this, we employ statistical analysis to explore if the precipitate chemistry varies with boundary type. This builds upon prior work by others, including analysis of variance (ANOVA) with SEM based image analysis by Jehadi *et. al.* [19], [20]. The approach involves particle labelling, from EDS spectra and image processing, grain boundary identification with EBSD, and then use of ANOVA to explore variations in precipitate chemistry with boundary type. This is used to compare two different Alloy 625 samples manufactured using wrought and powder metallurgy based methods, with a view to understanding their stress corrosion cracking performance and inform their use in nuclear reactor pressure vessels.

**Materials and Methods**

Alloy 625 is a high chromium content nickel super alloy with defined content for the alloying additions. The bulk composition of the material is seen in Table 1.

| Elements | Limiting Chemical Composition / wt% |
|---|---|
| Ni | 58.0 min |
| Cr | 20.0-23.0 |
| Fe | 5.0 max. |
| Mo | 8.0-10.0 |
| Nb | 3.15-4.15 |
| C | 0.10 max. |
| Mn | 0.50 max. |
| Si | 0.50 max. |
| P | 0.015 max. |
| S | 0.015 max. |
| Al | 0.40 max. |
| Ti | 0.40 max. |

Table 1: Standard composition of Inconel 625 as produced by Special Metals [21]

This study characterises two Alloy 625 materials:

1. A HIP material was formed from powder metallurgy and then hot isostatically pressed (HIP)
2. A wrought material, manufactured using a conventional wrought processing route

From each material, representative samples were sectioned and hot mounted in Bakelite and prepared for investigation through a standard metallographic route with a final polishing step of 0.04 µm OPS solution, pH balanced using $H_2O_2$ and water in the ratios 4:1:1.

Analytical Scanning Electron Microscopy (SEM) was conducted using a Quanta FEG 650 SEM equipped with the Bruker eFlashHD2 and a Quantax 60 $mm^2$ X-ray detector. Data was processed using Bruker ESPRIT v2.1.2.17832, and data exported to .h5 format prior to analysis within Matlab 9.5.0.944444 (R2018b).

EBSD patterns and energy dispersive X-ray spectra were collected using a ~10 nA probe current at 20 kV, using a working distance of 15 mm and 70° sample tilt. A total of 3 regions of interest per sample were collected, and the scanning set up is detailed in Table 2. The number of points and exposure time were constrained by instrument availability.

| Sample | Step Size / µm | No. Points | Exposure Time / ms |
|---|---|---|---|
| Methods Map (HIP'ed sample) | 0.14 | 525 x 349 | 28 |
| HIP (x3) | 0.21 | 1375 x 915 | 11 |
| Wrought (x2) | 0.21 | 1375 x 915 | 11 |
| Long Exposure Wrought (x1) | 0.21 | 1375 x 915 | 45 |

Table 2: Sample exposure time and size of each area of interest.

EDS spectra were captured with 2048 energy bins, up to 20 keV (the input accelerating voltage) and the detector has an energy resolution as specified by the manufacturer as 126 eV Mn Ka. The EBSD patterns were collected at a resolution of 320 x 240 pixels with the exception of one HIP area of interest (AOI) that was collected at 200 x 150 pixels. The patterns were indexed 'online' using ESPRIT 2.1 and they were all indexed against the FCC Ni-phase only.

EDS spectra and EBSD measured crystal orientations were exported, via the h5 file, for post processing.

**Post Processing and Correlative EDS and EBSD analysis**

Correlative analysis was performed using both the EBSD data and EDS data. The EBSD data was used to create a grain structure map, and specifically identify the grain boundaries and classify them as special when they have an orientation relationship which matches a $\Sigma 3^n$ coincident site lattice (CSL) type. The EDS data was used to identify high Mo and Nb containing carbide particles, and later perform quantiative EDS analysis with additional statistical testing.

To produce the EBSD derived data, the orientation data was imported into MTEX v5.2 [22] to create orientation maps inorder to determine grain boundary locations and characterise their type. Grain boundaries were calculated with a >5° cut off and the grain boundary network was created. As the grain boundary map is determined with respect to the X-Y raster of the EBSD map, the network was smoothed using a a spline filter in MTEX, where the grains were also filtered to remove non indexed points in the map and grow the grains near (non indexed) inclusions creating a complete grain boundary network. Finally, the edge of the map was removed from the grain boundary network.

Next, each segment of the grain boundary map was classified by boundary type. The grain boundary axis and angle was classified with a tolerance of 5° and all other boundaries were labelled as "random". At this point, the grain boundary network includes positions and character of each segment within a labelled network.

Carbide particles were identified through analysis of maps derived from EDS spectra. At each point, the (background corrected) EDS spectrum was windowed and summed at 2.293 keV (the primary transition for Mo-L$\alpha$ is 2.292 keV and the Nb-L$\alpha$ peak is at 2.169 keV, for 20 kV accelerating voltage) with an energy window of ±0.8 kV. The frequency distribution of the intensity within this summed part of the EDS energy spectra is shown in Figure 1a.

To improve detection of carbide particles, a 3 x 3 point spatial-median filter was applied to the map and then the map of total windowed X-ray counts was binarised through contrast normalisation and thresholding.This threshold is manually controlled, the selected value for the Mo+Nb-L$\alpha$ peak is highlighted in Figure 1, which also shows the effect of the image processing steps that have increased the contrast. In the pre-processed histogram the particles intensity and matrix intensity for this peak can be seen in intensity windows shown in Figure 1a

At this point, we have a map of points that have element concentration higher than the threshold labelled with values of 1, as demonstrated in Figure 1b. This process labels particles of high X-ray generation for the specific peak energy (here it corresponds to combined generation of Mo and Nb-L$\alpha$). Once these particles are identified, later steps can quantify the spectra.

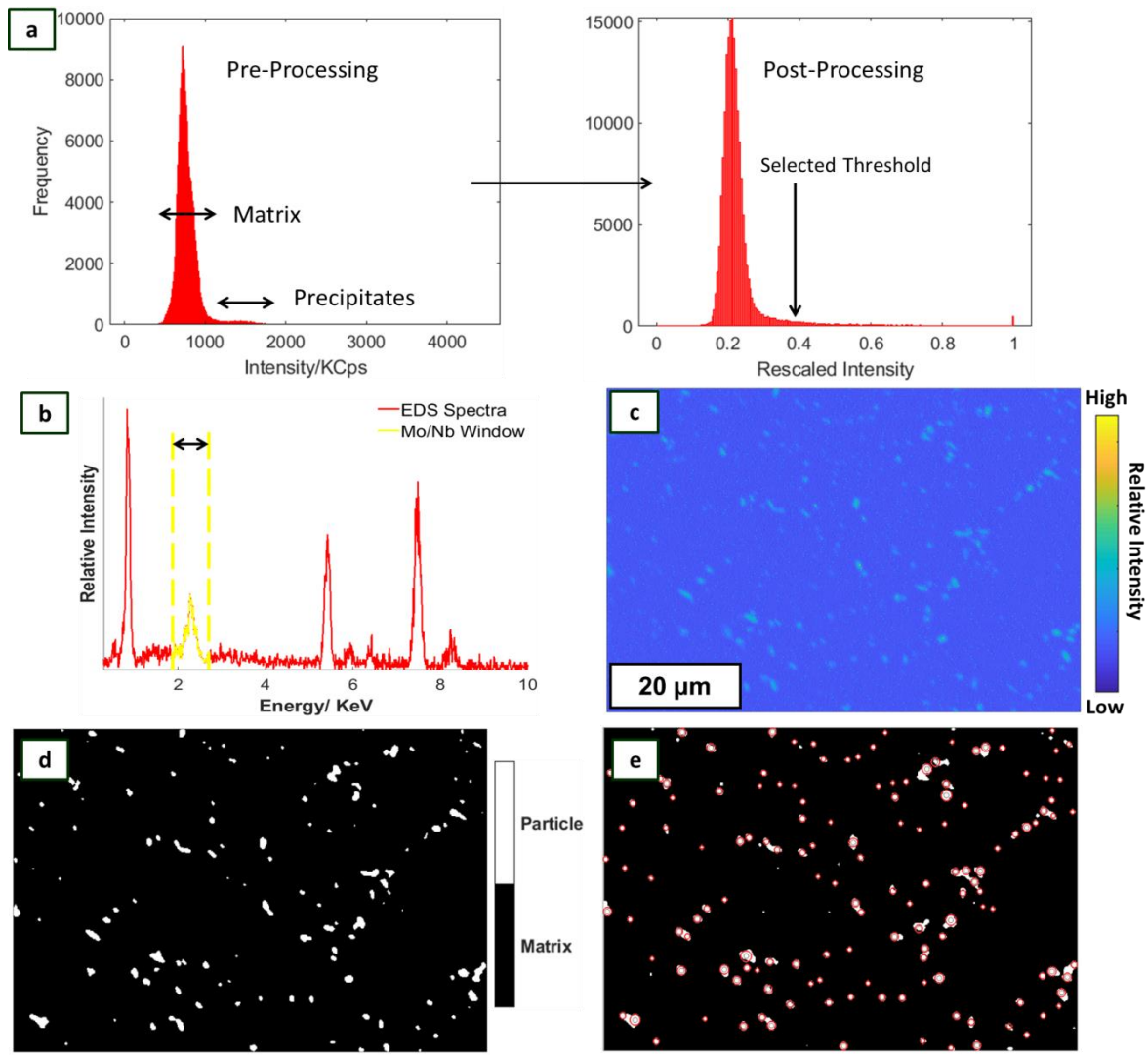

*Figure 1: The histograms for the pre and post processed Mo+Nb-Lα intensity map with the manually selected binarization threshold. The pre-processing histogram shows the two intensities of the precipitates and the matrix regions. A programmatic method for threshold selection was trialled. Labelling of carbides using EDS based analysis (b) EDS spectra from a single point. (c) Reconstructed spatial map from the summed intensity of the Mo+Nb-Lα yellow window from the energy window from each point. (d) Binary particle map resulting from noise filtering and thresholding the image. (e) Red circles show where a circular Hough transform has detected a particle in the binary image.*

Next, the particles were detected using a Hough based circle finding approach with the binarised map as the input data. The identified particles as identified are shown in Figure 1e. The circle finding algorithm was used to co-locate regions of high elemental concentration, to provide a centre of mass (in the 2D sectioning plane) and to quantify the distance of the particle from the nearest boundary (and the boundary type). For the purposes of this work the assumptions is made that all of the particles being investigated are spherical or near spherical.

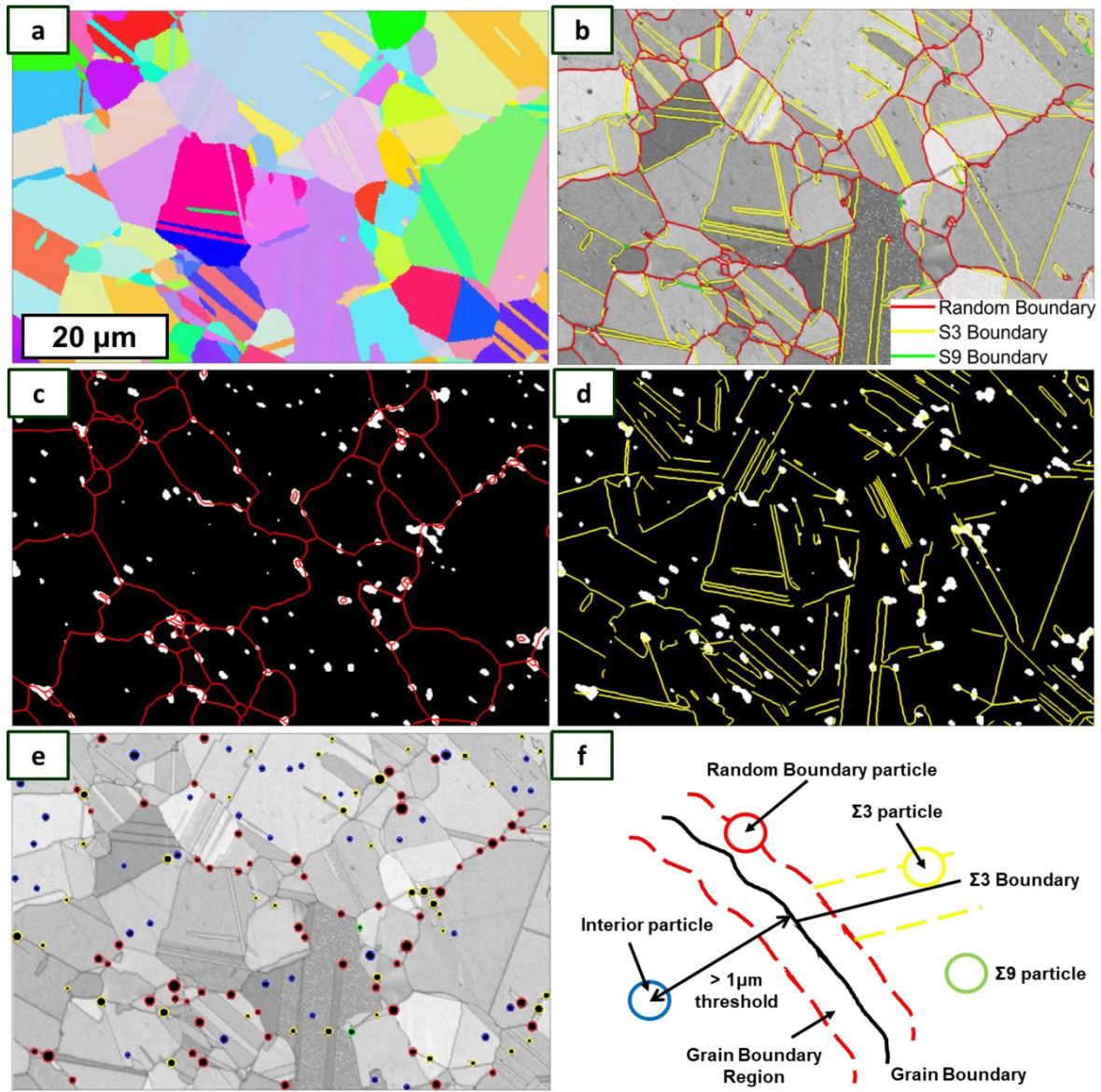

*Figure 2: Microstructure maps of HIP'ed Alloy 625 (a) crystal orientations. (b) Grain boundaries overlaid on a microstructure map, coloured by boundary type. The Σ3 and Σ9 are low energy 'special' CSL boundaries. Determining particle locations by combining EBSD and EDS data (c) Binary particle map with randomly orientated grain boundaries overlaid. (d) Binary particle map with Σ3 grain boundaries overlaid. (e) Microstructure map with identified particles separated and colour coded by location. Spectra are extracted from each point within each circle to enable comparison by precipitation location. (f) Cartoon of Euclidean distance calculation and threshold distance from circle to boundary for allocation of particle location. Distance is calculated from the central point of each circle.*

The location of each particle with respect to the boundary was determined by the Euclidian distance of the centre of the particle and the nearest grain boundary segment (obtained from the grain boundary network graph). The particle was considered to be on a specific boundary when the centre lies less than 1 μm from a boundary. This distance is five measurement points away from the boundary and is based upon the particle size distribution, as the average particle diameter identified was in the range of 1-2 μm. The result of particle identification and labelling is visualised in Figure 2e.

A threshold of 1 µm has been selected principally based upon the observed size and distribution of these particles in our materials. There are two additional aspects that we could consider which include: (1) we are assessing a 3D structure with a 2D section, and therefore the exact distance will depend on the size of the particle and the sub-surface grain boundary inclination; (2) the physics of the observation method, where the physical interaction volume of the EDS signal can be taken into account, which again suggests that a 1 µm threshold is reasonable (given the Z-numbers of the precipitates here and Monte Carlo based analysis of the volume).

Once the particles were identified, the sum EDS spectra for all points within the particle were extracted and reported back to within EPSRIT using an algorithm developed by T. McAuliffe *et.al.* [23].The ZAF based approach in EPSRIT is used to then quantify all of the exported spectra as we are more interested in accuracy for the heavier elements. ZAF is more accurate for the heavier elements while the addition of the ϕρ(z) method is more accurate for lighter elements [24].

**Reliability of Hough based Circle Identification**

Validation of the Hough based circle identification routine was performed using a simulation, and an example of this validation is presented in Figure 3. This helps address concerns regarding incomplete sampling and stochastic noise in the experimental data, analysis of individual circles is prone to small errors. The simulation was used to verify that the mean number of circles and their distribution was ensured, and that there was no introduction of sampling bias.

The simulated data was created using a random array of binarised circles in a discrete grid, some of which were overlapping. The size of the simulated particles and the discrete point-based sample of the experimental grid was used to inform the input distribution (the code for this simulation is included in the supplementary files).

Two separate approaches were implemented to reduce oversampling of features found within the experimental data. The artefacts introduced include deviations in circularity and overlapping particles, as identified from features seen in the binary images (Figure 3b). Due to the large imaging area involved, the use of a large input range for the radius in the 'imfindcircles' MATLAB function [25] was necessary to attempt to capture all available particles. This results in these smaller circles to be identified within larger features seen In Figure 3b, this is expected as the authors of the function suggest a small range in the desired radii [26]. For particles where there are multiple smaller circles, the first approach locates the largest circle in the group and removes any of the smaller circles that sit within in 1 diameter of this largest circle.

The second approach uses a barycentre method which enables the joining of overlapping smaller domains into one bigger circle. This replaces two over lapping circles with a weighted average to reposition a single circle of diameter equal to the larger of the two. This approach was used to address artefacts observed from visual inspections of the results.

A simulation was then used to evaluate quantitative artefacts. To determine if the radius can be used as a quantitative metric, the difference between the generated radii distribution and the detected distribution is identified. Figure 3c shows that the detected radii distribution is discrete as compared to the input (continuous) distribution, which is to be expected based upon the discrete sampling grid. The Kolmogorov-Smirnov test was used to confirm that 1) both are normally distributed and 2) the resulting p-values show to a 95% significance that the two distributions are equal. It is notable that as the total range of radii increases in the generated distribution, there is a trend in the reduction of the similarity of the two curves while still satisfying the KS test [27]. This could become a factor if the particle size range being investigated is large.

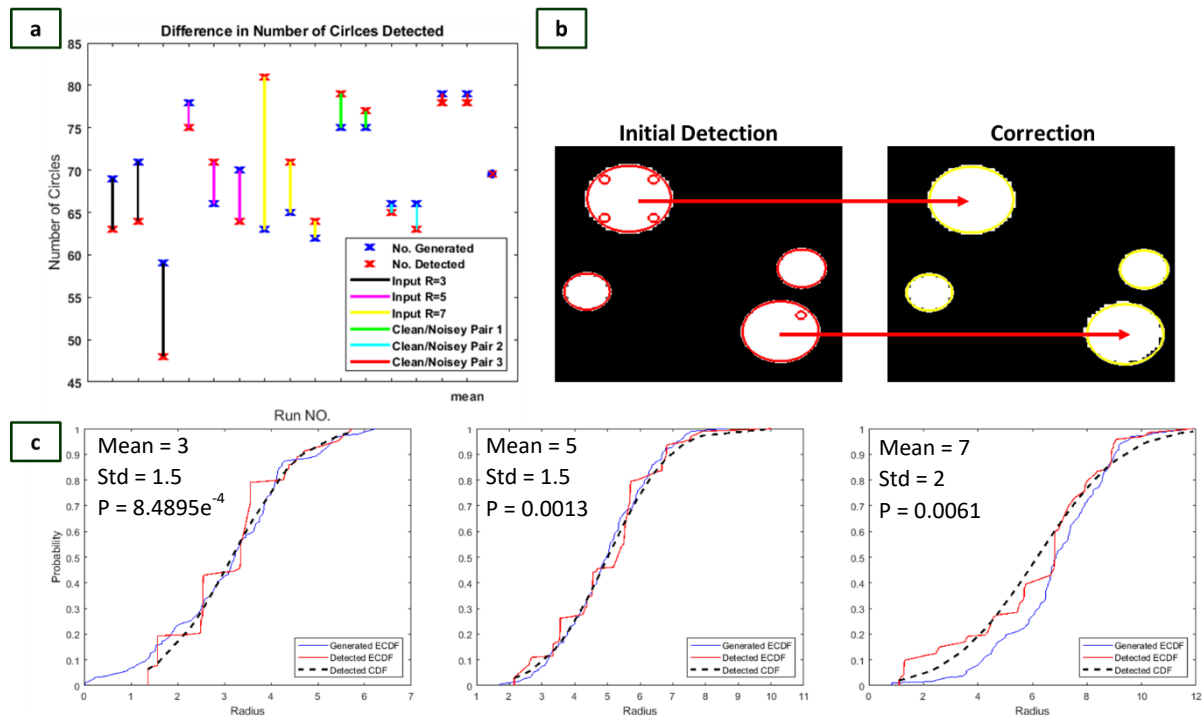

*Figure 3: Quantifying the effectiveness of the Hough based circle detection approach. (a) The variation in the number of circles generated and detected over a series of trials at different generated normal distributions. Including the impact of a layer of gaussian noise. (b) Two different corrections applied to the detected circles, a simple and barycentre approach to remove any excess incorrect labelling. (c) Comparing the empirical cumulative distribution functions (ECDF) for both generated and detected circle radii, with the normal CDF for the detected circles. The mean, standard deviation and result of a Kolmogorov-Smirnov test for similarity between the generated and detected ECDF's is shown for each case.*

As shown in Figure 3c, there is variation in the total number of particles detected in each individual case, which is to be expected, however importantly to this study the mean number of detected and generated circles are the same. This provides a level of confidence that, while in each individual experiment the identified number of features in the image may not be exact, the mean number of particles present should be within a reasonable range and representative of the sample.

Ultimately this simulation provides enough confidence that Hough transform based circle analysis method is suitable for sampling the precipitate distributions found in these samples. The Hough-transform based method was used on the experimental data to join together spectra associated with each particle, and to link these with the location and size of the particles to the EBSD derived grain boundary map.

**Effect of Specimen Tilt on Quantified EDS Analysis of Carbide Chemistry**

Now that we have the EDS data clustered and segmented for each individual particle, quantification can be performed. For surface investigation and microanalysis, the sample surface is typically positioned perpendicular to the beam direction to produce the smallest possible interaction volume which is as isotropic as possible (i.e. assuming the electron beam enters an infinite half space). However, when conducting simultaneous EBSD and EDS data collection with this method it is not possible as typically the sample must be tilted to 70° to allow for the maximum quantity of backscattered electrons to be incident on the EBSD detector. The EDS interaction volume is then consequently larger and anisotropic.

The impact of this tilting on the EDS signal was investigated by Monte-Carlo simulation using CASINO v2 [14]. Additionally, this enables us to consider where the Mo+Nb-Lα sum peak is suitable for particle shape reconstruction due to the difference in interaction volume and thickness variation on the intensity of the peaks.

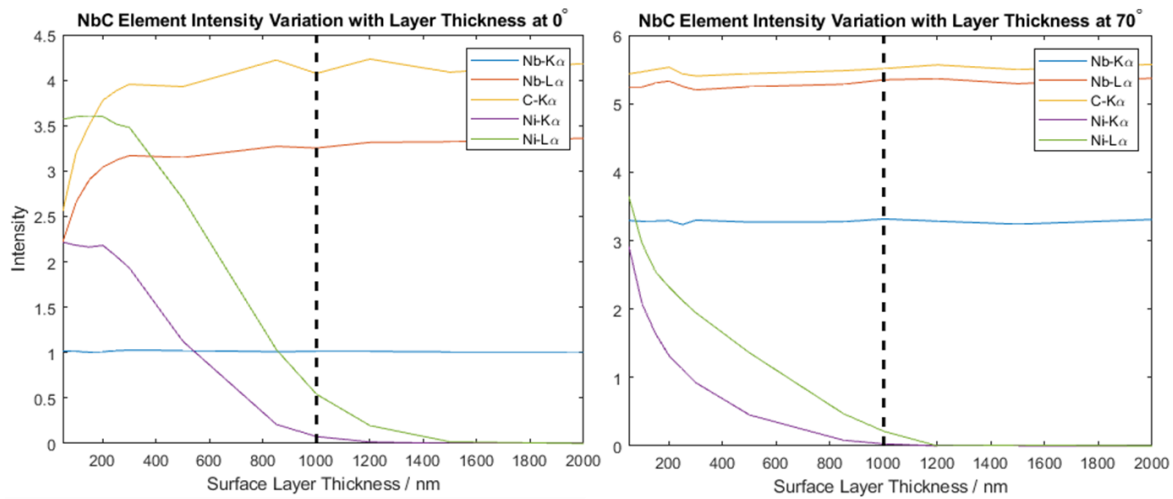

*Figure 4: The effect of sample tilt modelled by Monte-Carlo simulation with CASINO of increasing thickness of a surface of NbC with an infinite Ni layer beneath represented by taking the maximum generated X-ray intensity from the phi-rho-z curves.*

In the Monte Carlo simulation, ten thousand electron paths were simulated with an incident beam energy of 20 kV for both a 0 and 70° sample. In Figure 4, the solid black line shows the thickness at which for ZAF corrections based on the interaction volume of X-ray production (in the region of 1 µm), that the total interaction volume for any particular path is contained within the target. In this case that would be the NbC particle. Figure 4 shows a difference in the simulated maximum X-ray intensity generated of all transitions when the thickness is varied at both tilts however the maximum X-ray intensity of the Nb-Lα peak is effectively consistent at all particle thicknesses in a titled sample. There is a potential issue relating the maximum generated X-ray intensity taken from the phi-rho-z curves [28] as the absorption of X-rays requires and additional step to be accounted for. However due to the geometry of the samples, the path length of an X-ray in the tilted configuration is smaller and therefore the intensity loss from absorption will be smaller.

This simulation provides a reasonable indication that the intensity from the element seen in the surface layer will follow trends to those seen in existing phi-rho-z curves. For the quantification, and segmentation, here the simulation suggests that the shapes of particles reconstructed in the binary image are likely to be similar to those observed when sampling an un-tilted sample, as the measured intensity of X-rays from elements within the thresholding window are less thickness dependant. Furthermore, points at the edges of particles should be distinguishable from the matrix using the simple threshold.

Once the data is clustered into the sum peak, this reduces subsurface thickness variations and the ZAF correction within ESPRIT is used, which includes a tilt correction component to account for the absorption effect.

**Segmented EDS data analysis and precipitate distributions**

With the particle data classified by location, the spectra for each single category were collated and the mean particle spectra was created for each boundary type which can be seen in Figure 5a where they are compared to show that in the spectra there is a difference in intensity between Nb, Mo, Ni and possibly Cr. As these spectra are the mean values for all particles of each class, there is potential for these to be masking variation within one class. This motivates statistical analysis using Analysis of Variance (ANOVA) test to establish whether there are significant differences between the classes, as compared to the variance within each class.

The EDS quantified spectra are treated with a standard statistical approach to identify likelihood of similarity using ANOVA [29], [30]. Each boundary category contains the quantified data for elements identified through ESPRIT. In an ANOVA statistical test, the hypothesis states that the elements across each boundary will show no statistical difference, i.e. they concentration of an element is generally consistent on each boundary. For an ANVOA test there is an assumption that the data is normally distributed therefore, to validate the use of ANOVA, the KS test is used on each element for each boundary to determine if the composition can be described by a normal distribution. The validation metric in this test is also a P-value to a 95% significance. In the case of this example, the result of the KS test indicates that the distribution of concentration is normal suggesting that the differences highlighted, or not, by the ANOVA test are significant.

Figure 5b shows the quantified EDS spectra comparison between the boundaries highlighting similar patterns in concentration variation as indicated by the mean chemical spectra.

We observe a variation in concentration of Nb and Mo for the inclusions, and also Cr, by boundary type. This is confirmed in the ANOVA analysis shown for Nb in Figure 5c where the difference in concentration between the intragranular particles and those precipitating on the randomly orientated grain boundaries are, to a 95% significance, different indicating preferential location of Nb. The same analysis is repeated for Mo and Cr shows to the same significance that there is variation in concentration by particle location in the sample.

Using the ANOVA method, the P-value is used to indicate that it is likely that there is at least one boundary category that the concentration of that selected element is significantly different to at least one class of boundary. We note that the absolute P-value is not as important if the target significance threshold is met – in this case the significance is set to 95% [30]. To be completely certain of statistically identified trends, enough data would need to be collected that any addition of new data sets would produce the same identified concentration difference.

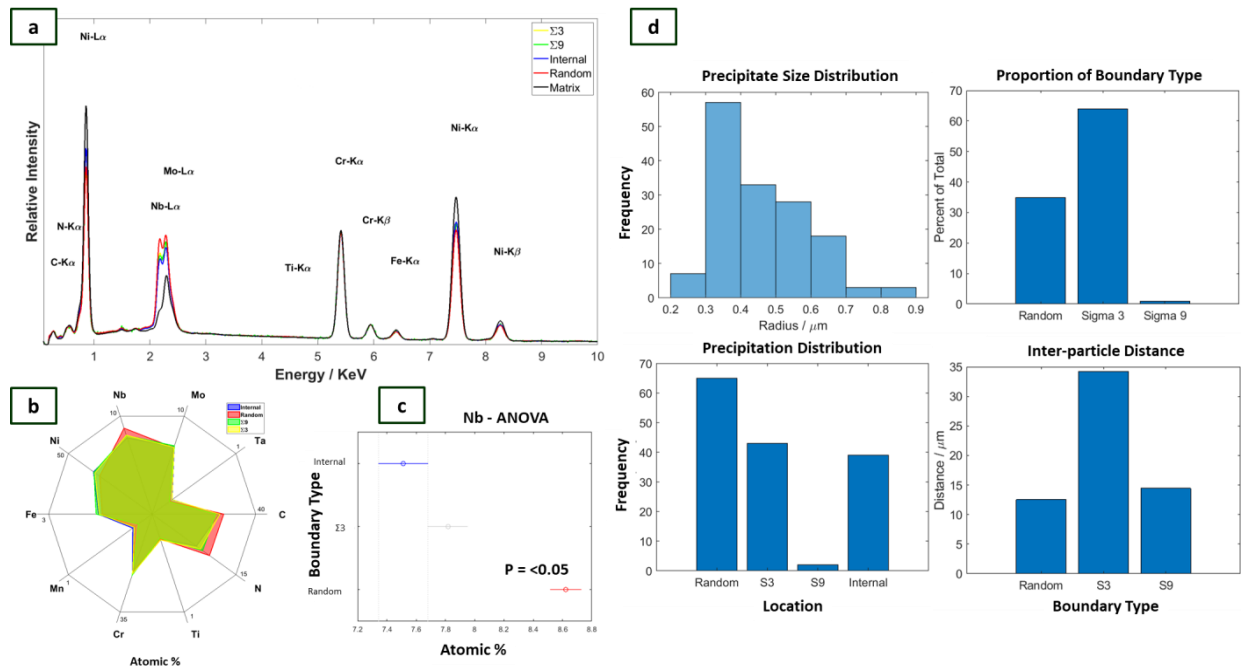

*Figure 5: Segmented EDS spectra by boundary category compared as raw spectra, quantified atomic % and statistically compared through ANOVA. (a) Mean EDS spectra across each boundary. (b) Radar plot of the atomic % of element present by boundary category from the marked particles. (c) Comparison of the calculated mean, with 95% confidence intervals for Nb content, compared by boundary type. (d)The quantification metrics for precipitate distributions in size range, location, boundary content and particle distribution along each boundary.*

As there is statistically significant variation in chemical composition of the particles by microstructural location, it is also possible to identify if there is preferential locations for these precipitates of different composition. The number of identified particles is highlighted in Figure 5c to show that there is a preference of Nb, Mo rich precipitates to locate on the randomly orientated boundaries with the largest number of particles and the smallest inter particle spacing located along this boundary type.

These metrics, as seen in Figure 5, have been chosen as they are likely important for corrosion and stress corrosion cracking studies while also informing more general material characterisation.

**Application to a Case Study in the Effect of Processing Route on Resultant Microstructure**

The difference in the resultant microstructure between two different processing routes, powder metallurgy hot isostatic press (HIP) and wrought (WR), of Alloy 625 are now investigated. The EBSD-EDS mapping details are presented in Table 2.

A sample of each microstructure is shown, with the binarized element map combined with EBSD derived grain boundary structure for each case in Figure 6.

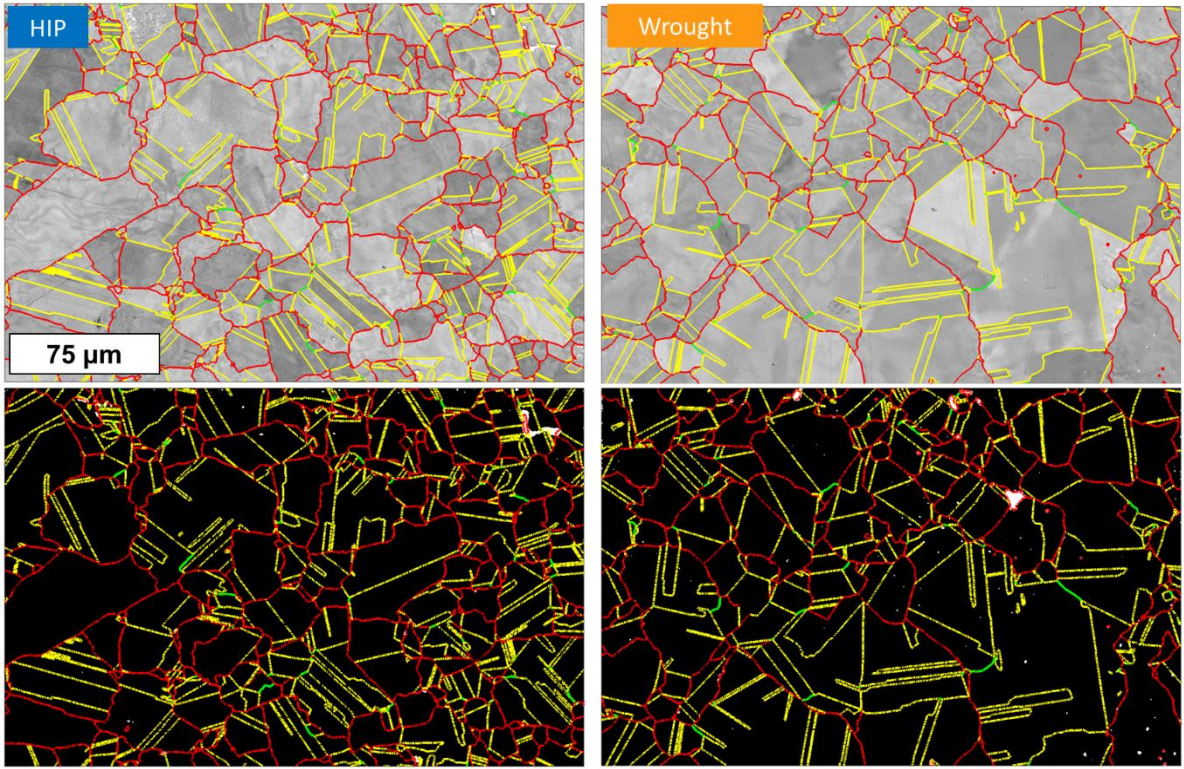

*Figure 6: The microstructure of a sample of HIP and Wrought microstructures (top) with the accompanying binary particle map and overlaid grain boundary structure (bottom).*

Each map is processed as per the method described above for this case study. Sum spectra for each particle type within each manufacturing route were analysed with results shown in Figure 7.

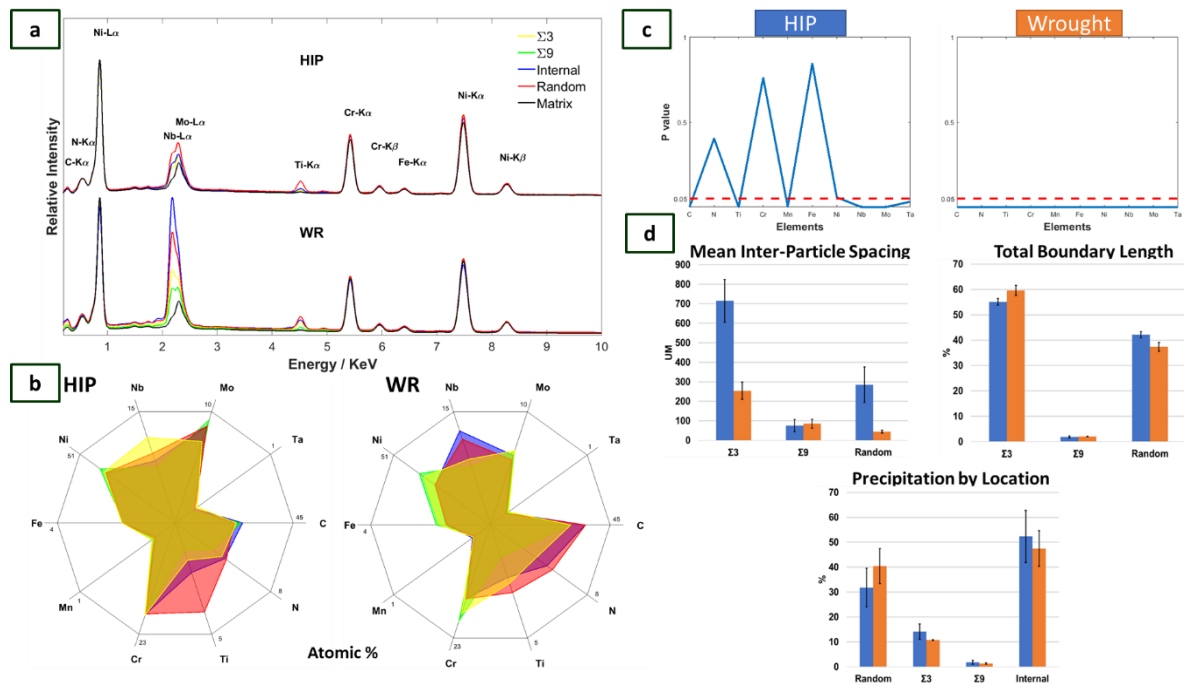

*Figure 7: Quantitative particle and boundary analysis comparing the HIP and wrought (WR) microstructures. (a) The mean chemical spectra for particles identified along each boundary highlighting the difference in the particle compositions and the similarity in the matrix compositions. (b) Quantified mean spectra in Atomic %. (c) Resulting P-Values of ANOVA for each individual element compared between each segment. (d) Proportion of the total grain boundary length that is each of Σ3, Σ9 (special boundaries) and randomly misoriented grain boundaries. The proportion of total identified particle compared by precipitation location identified and the mean spacing of the particles along each of the boundary types.*

Figure 7 shows there is a statistically significant difference in the boundary content of some of the elements within the HIP microstructure. Figure 7b shows distinct composition differences of the precipitates by processing route and 7d shows that the distribution of particles within each microstructure is different. For Nb in the HIP material the highest concentration is found in the precipitates laying along the Σ3 boundaries while in the wrought material this is found in intra-granular particles.

The inter-particle spacing along each of the boundaries also varies greatly due to the number of particles precipitating; and there are far fewer particles identified in the HIP microstructure compared to wrought. While the proportion of boundary length is very similar, there is a slightly higher special boundary (Σ3 and Σ9) content present in the wrought sample. There is a larger standard error in the precipitation by location, however there is still evidence that the distribution of precipitation is likely to remain distinct between processing routes.

**Discussion**

The use of alloy 625 widely deployed in high temperature applications within a nuclear reactor which has an environment the leads to Stress Corrosion Cracking (SCC). This motivates understanding of microstructural variations that can influence SCC resistance. These are broadly introduced by Rebak and Szklarska-Smialowska [31]. This highlights that SCC resistance can be influenced by microstructural effects (as well as the operating environment). Broadly in SCC and microstructural analysis, many either study the proportion of special boundaries or they look at the variation of chemistry and precipitates. This motivated our current study where we combine the two to see if they are related.

Hu *et al.* [32] identify that the proportion of CSL grain boundaries present in a microstructure can impact the rate and amount of SCC that affects the material. This is also explored by Bruemmer *et al.* [33] who quantify precipitate distributions and suggest understanding of these features could enable the prediction of SCC behaviour of the material. As was found in the case study, both of these factors have been quantified between the two processing routes and shown to have identifiable differences suggesting differences in the resulting SCC performance.

Our results here indicate that the chemistry and distribution of particles is a function of boundary type (or vice versa) which indicates that a correlative method is required for us to understand the influence of microstructure on SCC performance. This can vary between manufacturing routes.

For our correlative approach using simultaneous EBSD and EDS, the main limiting factor in what particles types are detectable is the interaction volume on the tilted surface. For EDS, the interaction volume is considered to be roughly 1μm in diameter. A step size of 0.21 μm was chosen to include precipitation in the size range of the EDS interaction volume. As a result, it is likely that precipitates in the finer size range of less than 1 step size (0.21 μm) will be missed from this data collection method. As we have not used the diffraction patterns in this study we are also unable to identify specific phases that the chemistry is contributing to, only that regions of high concentration of a particular element exist and that the statistics highlight that this is varied depending on precipitation location. Due to this step size, it is understood that smaller phases that are seen to precipitate in Alloy 625 that are rich in Cr will be missed by this current method of analysis which presents the opportunity for further investigation using higher spatial resolution methods.

There is significant use of image binarization and boundary detection for particle size analysis in the field of geology which employ the more traditional method of particle detection [34]–[36] and in commercial products such as ImageJ. However, for this study the circular Hough transform method provided a simpler approach within the Matlab based analysis framework to enable the further analysis for the combined processing shown here. The Circular Hough transform method has seen extensive use within the field of TEM particle counting/ radius determination and in material science [37]–[40] as well as similar application within general particle analysis, for example in the case of counting the number of cells [41], [42] or identifying particles in a flow [43] There are studies that similarly identify that the radius determined from the circular Hough transform is at least reliable for the image [37], how this then relates the 3d shape of particles has not been determined.

Within the Alloy 625 system, the precipitates highlighted by this method, the Mo, Nb rich particles are of FCC type crystal structures with very similar lattice parameters to the Ni matrix. This presents a difficulty in distinguishing these particles from the matrix with the conventional EBSD method used in this work [44], [45]. There have been methods developed that could use the Kikuchi diffraction data to enable the identification of phases when there is only small differences in the patterns by Foden *et al.* [46].

For the best statistics, a sufficiently large area should be studied with a reasonable sampling resolution. To address this challenge in the present work, three randomly selected different areas of the surface for each sample were measured to create an approximation to the total microstructure of the samples. However it should be identified that more data could provide further insight.

As we now have quantitative mapping of particles linked to boundary types for the different product forms, we can move beyond anecdote and use statistical testing. In terms of the application of

ANOVA, this approach has been applied to statistics of EDS data in previous studies to quantify differences in chemical content [20].

A next step might involve application of direct linking of the EBSD and EDS data, e.g. using multivariate statistics has been used singularly to perform image segmentation [47] or in combination with EBSD as in McAuliffe *et al.* [23]*,* and this could assist in identification of the precipitates. However each of these methods in isolation would not provide the type of information desired to relate to SCC.

To complement our present work, high spatial resolution analysis is also possible, such as with Atom Probe [48], TEM [49] or STEM [50], which are extremely effective in determining the phase, chemistry and location of precipitates at the fine scale. However each of these cases requires significant time investment in sample preparation and are only really possible to conduct on very small volumes containing a small number of precipitates before conclusions are drawn about the material on larger scale. This means there is limited opportunity for these techniques to look at a large population of precipitates, such as the Mo and Nb rich precipitates sampled here, and to explore the statistical distributions more widely. For larger more scale investigations of precipitation, synchrotron based investigations are generally conducted [51]–[53], e.g. with small angle X-ray scattering and/or diffraction, but spatial correlation is lost.

The present work is prompting future work in the application of this method to TKD and EDS to allow for accessing relatively large areas of interest to create larger data sets allowing more accurate statistics. This would enable the investigation into whether similar patterns appear in the fine scale precipitation and to determine if there is the same difference in concentration by location. Investigating the material in both length scales can provide insights into how the material is going to behave and what the full impacts of the manufacturing processes are.

**Conclusions**

A method has been developed to use simultaneously collected EBSD and EDS data to provide information on the chemistry and distribution of precipitates in a Ni-rich matrix and their correlations with different grain boundary types. Statistical analysis, using ANOVA, highlights that there are variations in the chemistry between precipitates found on different boundary types as well as between samples manufactured using two different processing routes for Alloy 625. This may support further studies to understand the stress corrosion cracking performance of Alloy 625.

**Acknowledgements**

We acknowledge the support of Tom McAuliffe in exporting the EDS data from Matlab to ESPRIT . We acknowledge funding from the EPSRC Centre for Doctoral Training in Nuclear Energy (EP/L015900/1) and Rolls Royce Submarines Limited for CB's studentship. TBB acknowledges funding of his research fellowship from the Royal Academy of Engineering. Electron microscopy was conducted within the Harvey Flower EM suite on an instrument funded by the Shell-Imperial AIMS UTC.

**Data Statement**

The data can be downloaded from Zenodo (10.5281/zenodo.4507808).

**Author Contributions**

CB collected and analysed the data, developed the analysis routines and wrote the first draft of the manuscript. AB and TBB supervised the work. All authors contributed to the final manuscript.

**References**

[1] P. Burdet, S. A. Croxall, and P. A. Midgley, "Enhanced quantification for 3D SEM-EDS: Using the full set of available X-ray lines," *Ultramicroscopy*, vol. 148, pp. 158–167, 2015.

[2] M. M. Nowell and S. I. Wright, "Phase differentiation via combined EBSD and XEDS," *J. Microsc.*, vol. 213, no. 3, pp. 296–305, Feb. 2004.

[3] A. A. Salem, M. G. Glavicic, and S. L. Semiatin, "A coupled EBSD/EDS method to determine the primary- and secondary-alpha textures in titanium alloys with duplex microstructures," *Mater. Sci. Eng. A*, vol. 494, no. 1–2, pp. 350–359, Oct. 2008.

[4] J. H. Kim, D. I. Kim, J. H. Shim, and K. W. Yi, "Investigation into the high temperature oxidation of Cu-bearing austenitic stainless steel using simultaneous electron backscatter diffraction-energy dispersive spectroscopy analysis," *Corros. Sci.*, vol. 77, pp. 397–402, 2013.

[5] G. D. West and R. C. Thomson, "Combined EBSD / EDS tomography in a dual-beam FIB / FEG – SEM," vol. 233, no. May 2008, pp. 442–450, 2009.

[6] J. Goulden, S. Sitzman, K. Larsen, and R. Jones, "Real-Time Discrimination of Phases with Similar Kikuchi patterns but Different Chemistry through Simultaneous EBSD and EDS," *Microsc. Microanal*, vol. 21, no. 3, p. 2015, 2019.

[7] F. Cruz-Gandarilla, O. Coreño-Alonso, J. G. Cabañas-Moreno, L. H. Chan, C. Gómez-Esparza, and R. Martínez Sánchez, "Simultaneous XEDS-EBSD Study of NiCoAlFeCu(Cr,Ti) Multi-Component Alloys," *Adv. Eng. Mater.*, vol. 1700215, pp. 1–8, 2017.

[8] C. L. Chen and R. C. Thomson, "The combined use of EBSD and EDX analyses for the identification of complex intermetallic phases in multicomponent Al-Si piston alloys," *J. Alloys Compd.*, vol. 490, no. 1–2, pp. 293–300, Feb. 2010.

[9] S. Zaefferer and P. Romano, "Attempt to identify and quantify microstructural constituents in low-alloyed TRIP steels by simultaneous EBSD and EDS measurements," *Microsc Microanal*, vol. 13, no. 2, 2019.

[10] K. Moran and R. Wuhrer, "X-ray mapping and interpretation of scatter diagrams," *Microchim. Acta*, vol. 155, no. 1–2, pp. 209–217, 2006.

[11] R. Wuhrer and K. Moran, "Quantitative X-ray mapping, scatter diagrams and the generation of correction maps to obtain more information about your material," *IOP Conf. Ser. Mater. Sci. Eng.*, vol. 55, no. 1, 2014.

[12] J. I. Goldstein, D. E. Newbury, J. R. Michael, N. W. M. Ritchie, J. H. J. Scott, and D. C. Joy, *Scanning Electron Microscopy and X-Ray Microanalysis*, 4th ed. Springer, 2018.

[13] A. Deal, X. Tao, and A. Eades, "EBSD geometry in the SEM: Simulation and representation," in *Surface and Interface Analysis*, 2005, vol. 37, no. 11, pp. 1017–1020.

[14] D. Drouin, A. R. Coutre, D. Joly, X. Tastet, V. Aimez, and R. Gauvin, "CASINO V2.42 - A Fast and Easy-to-use Modeling Tool for Scanning Electron Microscopy and Microanalysis Users," *Scanning*, vol. 29, pp. 92–101, 2007.

[15] J. Crawford, D. Cohen, G. Doherty, and A. Atanacio, "Calculated K, L and M-shell X-ray line intensities for light ion impact on selected targets from Z=6 to 100," 2011.

[16] M. Nuspl, W. Wegscheider, J. Angeli, W. Posch, and M. Mayr, "Qualitative and quantitative determination of micro-inclusions by automated SEM/EDX analysis," *Anal. Bioanal. Chem.*, vol. 379, no. 4, pp. 640–645, 2004.

[17] K. F. J. Heinrich, C. E. Fiori, and R. L. Myklebust, "Relative transition probabilities for the x-ray lines from the K level," *J. Appl. Phys.*, vol. 50, no. 9, pp. 5589–5591, 1979.

[18] A. J. Wilkinson and T. B. Britton, "Strains, planes, and EBSD in materials science," *Mater. Today*, vol. 15, no. 9, pp. 366–376, Sep. 2012.

[19] M. Jahedi, E. Ardjmand, and M. Knezevic, "Microstructure metrics for quantitative assessment of particle size and dispersion: Application to metal-matrix composites," *Powder Technol.*, vol. 311, pp. 226–238, Apr. 2017.

[20] L. Moretti, M. Conficconi, S. Natali, and A. D'Andrea, "Statistical analyses of SEM-EDS results to predict the quantity of added quicklime in a treated clayey soil," *Constr. Build. Mater.*, vol. 253, 2020.

[21] Special Metals, *Inconel Alloy 625*, vol. 625, no. 2. 2013.

[22] G. Nolze and R. Hielscher, "Orientations - Perfectly colored," *J. Appl. Crystallogr.*, vol. 49, no. 5, pp. 1786–1802, 2016.

[23] T. P. McAuliffe, A. Foden, C. Bilsland, D. Daskalaki-Mountanou, D. Dye, and T. B. Britton, "Advancing characterisation with statistics from correlative electron diffraction and X-ray spectroscopy, in the scanning electron microscope," vol. 211, no. August 2019, 2019.

[24] J. J. Friel, *X-ray and Image Analysis in Electron Microscopy*, Second. Princeton, 2003.

[25] "imfindcircles." [Online]. Available: https://uk.mathworks.com/help/images/ref/imfindcircles.html.

[26] The MathWorks Inc., "Matlab v9.5.0.94444 (R2018b)."

[27] J. . Frank J . Massey, "The Kolmogorov-Smirnov Test for Goodness of Fit," *J. Am. Stat. Assoc.*, vol. 46, no. 253, pp. 68–78, 2017.

[28] G. Love, M. G. C. Cox, and V. D. Scott, "A simple Monte Carlo method for simulating electron-solid interactions and its application to electron probe microanalysis," 1977.

[29] L. S. Kao and C. E. Green, "Analysis of Variance: Is There a Difference in Means and What Does It Mean?," *J. Surg. Res.*, vol. 144, no. 1, pp. 158–170, 2008.

[30] L. Stahle and S. Wold, "Analysis of Variance," *Chemom. Intell. Labratory Syst.*, vol. 6, pp. 259–272, 1989.

[31] R. B. Rebak and Z. Szklarska-Smialowska, "The Mechanism of Stress Corrosion Cracking of Alloy 600 in High Temperature Water," *Corros. Sci.*, vol. 38, no. 6, pp. 971–988, 1996.

[32] H. U. Hong, B. S. Rho, and S. W. Nam, "Correlation of the M23C6 precipitation morphology with grain boundary characteristics in austenitic stainless steel," *Mater. Sci. Eng. A*, vol. 318, no. 1–2, pp. 285–292, 2001.

[33] S. M. Bruemmer and G. S. Was, "Microstructural and microchemical mechanisms controlling intergranular stress corrosion cracking in light-water-reactor systems," *J. Nucl. Mater.*, vol. 216, pp. 348–363, Oct. 1994.

[34] G. H. A. J. J. Kumara, K. Hayano, and K. Ogiwara, "Image analysis techniques on evaluation of particle size distribution of gravel," *Int. J. GEOMATE*, vol. 3, no. 1, pp. 290–297, 2012.

[35] C. W. Liao, J. H. Yu, and Y. S. Tarng, "On-line full scan inspection of particle size and shape using digital image processing," *Particuology*, vol. 8, no. 3, pp. 286–292, Jun. 2010.

[36] M. Damadipour, M. Nazarpour, and M. T. Alami, "Evaluation of Particle Size Distribution Using an Efficient Approach Based on Image Processing Techniques," *Iran. J. Sci. Technol. - Trans. Civ. Eng.*, vol. 43, no. 1, pp. 429–441, Jul. 2019.

[37] Y. Meng, Z. Zhang, H. Yin, and T. Ma, "Automatic detection of particle size distribution by image analysis based on local adaptive canny edge detection and modified circular Hough transform," *Micron*, vol. 106, pp. 34–41, Mar. 2018.

[38] S. Kook *et al.*, "Automated Detection of Primary Particles from Transmission Electron Microscope (TEM) Images of Soot Aggregates in Diesel Engine Environments," *Int. J. Engines*, vol. 9, no. 1, pp. 279–296, 2016.

[39] M. Mirzaei and H. K. Rafsanjani, "An automatic algorithm for determination of the nanoparticles from TEM images using circular hough transform," *Micron*, vol. 96, pp. 86–95, May 2017.

[40] S. Bourrous *et al.*, "A semi-automatic analysis tool for the determination of primary particle size, overlap coefficient and specific surface area of nanoparticles aggregates," *J. Aerosol Sci.*, vol. 126, pp. 122–132, Dec. 2018.

[41] S. Pavithra MPhil Scholar and J. Bagyamani Associate Professor, "White Blood Cell Analysis Using Watershed and Circular Hough Transform Technique," *Int. J. Comput. Intell. Informatics*, vol. 5, no. 2, 2015.

[42] S. N. M. Safuan, R. Tomari, W. N. W. Zakaria, and N. Othman, "White blood cell counting analysis of blood smear images using various segmentation strategies," in *AIP Conference Proceedings*, 2017, vol. 1883.

[43] J. Tan and P. Jin, "Three-dimensional particle image tracking for dilute particle-liquid flows in a pipe," *Meas. Sci. Technol.*, vol. 18, no. 6, 2007.

[44] S. Floreen, G. E. Fuchs, and W. J. Yang, "The Metallurgy of Alloy 625," in *Superalloys 718, 625, 706 and Various Derivatives*, 1994, pp. 13–37.

[45] D. Verdi, M. A. Garrido, C. J. Múnez, and P. Poza, "Cr3C2 incorporation into an Inconel 625 laser cladded coating: Effects on matrix microstructure, mechanical properties and local scratch resistance," *Mater. Des.*, vol. 67, pp. 20–27, Feb. 2015.

[46] A. Foden, D. M. Collins, A. J. Wilkinson, and T. B. Britton, "Indexing electron backscatter diffraction patterns with a refined template matching approach," *Ultramicroscopy*, vol. 207, Dec. 2019.

[47] C. Teng and R. Gauvin, "Multivariate Statistical Analysis on a SEM/EDS Phase Map of Rare Earth Minerals," *Scanning*, vol. 2020, 2020.

[48] D. Vaumousse, A. Cerezo, and P. J. Warren, "A procedure for quantification of precipitate microstructures from three-dimensional atom probe data," *Ultramicroscopy*, vol. 95, no. SUPPL., pp. 215–221, 2003.

[49] B. Hudson, J. P. Jackson, and I. Codd, "The effect of neutron irradiation on the precipitate distribution in adjusted uranium," 1974.

[50] D. Acevedo-Reyes, M. Perez, C. Verdu, A. Bogner, and T. Epicier, "Characterization of precipitates size distribution: Validation of low-voltage STEM," *J. Microsc.*, vol. 232, no. 1, pp. 112–122, Oct. 2008.

[51] W. Seifert *et al.*, "Synchrotron-based investigation of iron precipitation in multicrystalline silicon," *Superlattices Microstruct.*, vol. 45, no. 4–5, pp. 168–176, 2009.

[52] S. A. McHugo *et al.*, "Synchrotron-based impurity mapping," *J. Cryst. Growth*, vol. 210, no. 1, pp. 395–400, 2000.

[53] A. Steuwer *et al.*, "A combined approach to microstructure mapping of an Al-Li AA2199 friction stir weld," *Acta Mater.*, vol. 59, no. 8, pp. 3002–3011, 2011.